\documentclass[prl,onecolumn,showpacs,amsmath]{revtex4}

\usepackage{graphicx}
\usepackage[dvips]{color}

\begin{document}
\title{The nodal gap component as a good candidate for the superconducting order parameter in cuprates}

\author{W. Guyard, A. Sacuto, M. Cazayous, Y. Gallais and M. Le Tacon}
\affiliation{Laboratoire Mat\'eriaux et Ph\'enom$\grave{e}$nes Quantiques (UMR 7162 CNRS),
Universit\'e Paris Diderot-Paris 7, Bat. Condorcet, 10 rue Alice Domon et Léonie Duquet,75205 Paris CEDEX 13, France}

\author{D. Colson and A. Forget}
\affiliation{Service de Physique de l'Etat Condens\'{e}, CEA-Saclay, 91191 Gif-sur-Yvette, France}

\date{\today}

\begin{abstract}
Although more than twenty years have passed since the discovery of high temperature cuprate superconductivity, the identification of the superconducting order parameter is still under debate. Here, we show that the nodal gap component is the best candidate for the superconducting order parameter. It scales with the critical temperature $T_c$ over a wide doping range and displays a significant temperature dependence below $T_c$ in both the underdoped and the overdoped regimes of the phase diagram. In contrast, the antinodal gap component does not scale with $T_c$ in the underdoped side and appears to be controlled by the pseudogap amplitude. Our experiments establish the existence of two distinct gaps in the underdoped cuprates.

\end{abstract}


\maketitle
\date{\today}

In cuprate superconductors the critical temperature $T_c$ is strongly dependent on the carrier concentration (the doping level, \textit{p}) and exhibits a domelike shape with two distinct regimes (overdoped and underdoped). It is now established that the superconducting gap has a dominant $d$-wave symmetry across the entire superconducting dome [\onlinecite{Tsuei_PRL04}]. The superconducting gap reaches its maximum value along the antinodal directions and vanishes along the nodal directions corresponding respectively to the principal axes and the diagonal of the Brillouin zone. 
\par
Recent experiments have revealed a strong dichotomy between the nodal and the antinodal components of the gap in the superconducting state [\onlinecite{LeTacon_NaturePhysics06,Tanaka_Science06,Hanaguri_Natphys07,Hashimoto_PRB07}]. While the gap amplitude close to the nodes ($\Delta_{N}$) continuously tracks $T_c$, the antinodal one ($\Delta_{AN}$) is no longer proportional to $T_c$ and its  $2\Delta_{AN}/k_BT_c$ ratio blows up in the underdoped regime. A compilation of the experimental data (electronic Raman scattering (ERS), angle resolved photoemission (ARPES), neutron, tunneling and heat conductivity) which reveals this dichotomy is reported elsewhere [\onlinecite{Hufner_condmat07}].

\par
These results raise the question of "one or two gaps " in the cuprates [\onlinecite{Loram_JS1994,Tallon_PhysicaC2001,Deutscher_Nature99,MillisScience07}]. In the two gaps scenario, a small gap (close to the nodes) is  related to superconductivity whereas a larger one (at the antinodes), called the pseudogap, corresponds to a phenomenon distinct from superconductivity and may compete with it.
\par
A set of recent experiments [\onlinecite{Guyard_PRB77,Lee_Nature07,kondo,terashima_PRL2007,Valla_Science06,Krodyuk_condmat08}] suggests that the antinodal gap is indeed disconnected from superconductivity in the underdoped regime. The antinodal gap does not scale with $T_c$, becomes very weakly temperature dependent and even slightly increases as the temperature reaches $T_c$. These observations are in contradiction with what is expected for a BCS-like behavior as observed deep in the overdoped regime. 
\par
Since the antinodal component fails (at least in the underdoped side) to reproduce the behaviour expected for a superconducting order parameter, many efforts have been recently devoted to probe the temperature dependence of the nodal component. This issue is the center of an intense debate as recent ARPES experiments have yielded contradictory results. One group argues that the superconducting gap is essentially temperature independent over the whole $k$-space including the nodal region [\onlinecite{Kanigel_condmat08}] while another group argues that the amplitude of the gap near the nodes shows a well-defined BCS temperature dependence [\onlinecite{Lee_Nature07}]. 

\par
A clear identification of the temperature dependence of the superconducting gap is therefore of paramount importance. The superconducting nodal component is a priori a good candidate to \textit{feel} superconductivity  because its amplitude scales with T$_c$ throughout the entire superconducting dome as demonstrated by Raman scattering experiments on a wide range of cuprates [\onlinecite{Guyard_PRB77,YG-EPL,Opel_PRB00}]. 

\par
Probing the nodal gap structure as a function of temperature is challenging precisely because close to the nodes the gap amplitude vanishes. Moreover, the nodal gap structure is expected to be very sensitive to any thermal excitations and/or disorder inducing pair breaking which makes difficult the extraction of the bare gap function.

\par
In spite of such experimental difficulties, we bring here the first Raman experimental evidences that the nodal gap amplitude is temperature dependent in the overdoped and the underdoped regimes. In both regimes the nodal gap energy decreases as the temperature is raised and disappears at $T_c$. In sharp contrast, the antinodal gap component exhibits a distinct temperature dependence as a function of the doping level. It is not detected at low doping level. It is non temperature dependent (or weakly increases as the temperature is raised)  below the optimal doping while in the overdoped regime it shows strong temperature dependence consistent with a conventional BCS behavior.

\par
 Our results favor a scenario in which the nodal gap component remains continuously connected to superconductivity as the doping changes while the antinodal component is only related to superconductivity in the overdoped regime. 
They imply the existence of two gaps in the superconducting state of underdoped cuprates in agreement with recent ARPES  [\onlinecite{Lee_Nature07, Krodyuk_condmat08}] and tunneling measurements [\onlinecite{Boyer_Natphys07,Gomes_Nature07,Hanaguri_Natphys07}].

\par

We have probed the nodal and antinodal components of the gap by performing an ERS study on $HgBa_{2}CuO_{4+\delta }$ (Hg-1201) single crystals as a function of both temperature and doping level (from $p=~0.25$ to ~$0.09$). Hg-1201 has one single CuO$_2$ layer by unit cell and its doping level is controled by insertion of oxygen atoms in HgO layer. The ERS experiments have been carried out using a triple grating spectrometer (JY-T64000) equipped with a nitrogen cooled CCD detector. The nodal ($B_{2g}$) and antinodal ($B_{1g}$) regions have been explored by using cross polarizations along the Cu-O bond directions and at 45$^o$ from them respectively [\onlinecite{sacuto_PRB00}]. The red ($1.9~eV$) excitation line was used to probe both the nodal and antinodal components of the gap. 
The reason is that we have experimentally found that the ERS nodal gap peaks are sharper when using the 1.9~eV line compared to higher energy lines. This is possibly due to a sharper ($B_{2g}$) Raman vertex function near the nodes.
All the Raman spectra presented here were corrected for the spectrometer response, the Bose factor and the optical constants and thus show the imaginary part of the Raman response function $\chi ^{\prime \prime}(\omega)$.

\par
In Fig.~1, are displayed the nodal ($B_{2g}$) and antinodal ($B_{1g}$) Raman response functions in the superconducting state substracted from the Raman responses just above $T_c$,  $\chi_{S} ^{\prime \prime}(\omega)-\chi_{N} ^{\prime \prime}(\omega)$  for two different doping levels $p = 0.11$ [(a) and (b)] , and $0.20$ [(c) and (d)] corresponding respectively to $T_{c}=75~K$ (underdoped) and $85~K$ (overdoped). The insets show the unsubstracted superconducting and normal response functions: $\chi_{S} ^{\prime \prime}(\omega)$ and $\chi_{N} ^{\prime \prime}(\omega)$. 
\par
In the overdoped case ($p=0.20$), Fig.~1(a) and 1(b), strong temperature dependences of the nodal and antinodal superconducting structures are observed. The $B_{2g}$ spectrum exhibits a peak close to $375~cm^{-1}$ at $T=10~K$ which softens in energy and decreases in intensity as the temperature raises before disappearing at $T_c$. Simultaneously, the low energy continuum (the negative part of the spectrum) fills in agreement with the transfer of spectral weight usually detected in the Raman spectra of high $T_c$ cuprates [\onlinecite{Devereaux_RMP07,Guyard_PRB77}]. The description is quite similar for the antinodal Raman response except that the antinodal gap is higher in energy ($405~cm^{-1}$) than the nodal one as expected for a $d$-wave gap. We detect a similar energy softening between $10~K$ and $70 ~K$ of about $25 \pm 5~cm^{-1}$  for both the nodal and antinodal components. 
\par

In the the underdoped case ($p=0.11$), Fig.~1(c) and 1(d), the most salient feature is that the nodal and antinodal Raman responses become drastically distinct. The nodal component is temperature dependent and shows an energy softening of about $25 \pm 5~cm^{-1}$ similar to the one in the overdoped regime. In contrast the antinodal response is flat with no trace of electronic superconducting component. The antinodal Raman responses do not show any difference up to $1000 ~cm^{-1}$ (see inset of Fig.~1(d)) [\onlinecite{note1}]. This confirms our previous results [\onlinecite{YG-PRB,LeTacon_NaturePhysics06}] which revealed that the antinodal Raman response is reduced before disappearing completely below $p=0.12$.
\par
The temperature dependence of the nodal component for several doping levels (two underdoped $p=0.10$ and $p=0.11$ and one overdoped $p=0.20$) is reported in Fig.~2. The temperature dependence of the antinodal component at $p=0.13$ is also shown for comparison.  We observe that even at relatively low doping level ($p=0.10$) the nodal component is temperature dependent with an energy softening of about $20 \pm 5 ~cm^{-1}$ whereas the antinodal gap component at $p=0.13$ displays a slight upward shift as $T_c$ is approached. 
However we can notice that the temperature dependence of the nodal component for the three doping levels does not follow the evolution expected for a $d-$wave BCS gap [\onlinecite{Carbotte_PRB95}] (see Fig.~2). One of the main reason is that contrary to ARPES which probes quasi-particles in a single direction of momentum space, ERS spectroscopy probes a small but finite region around the nodes whose extension is essentially controlled by the shape of the $B_{2g}$ Raman vertex function. 
We therefore expect (especially for underdoped or slightly overdoped measurements) the pseudogap which develops at the antinodes to contaminate the temperature dependence of the $B_{2g}$ Raman response function around the nodes. An estimation of the pseudogap component deduced from the temperature evolution of the antinodal component as a function of doping level will be adressed just below. It will appear that the pseudogap component (measured close to $T=T_c$) is still sizeable in the sligtly overdoped side and only disappears in the strongly overdoped regime.  Other causes could also be invoked such as disorder which induces pair breaking and produces a thermally activated low energy quasi-particle peak. Such a peak is expected to broaden with temperature and can mask the temperature dependendence of the nodal component just below $T_c$.
 
 \par
In order to quantify the pseudogap component as a function of the doping level, we have re-analysed our recent data [\onlinecite{Guyard_PRB77}] (see insert of Fig.~3) using a very simple model proposed by Loram et al. [\onlinecite{Loram_JS1994}].  We have postulated that the antinodal component $\Delta_{AN}$ has two contributions: the bare superconducting gap $\Delta_S$ which dominates in the overdoped side and the pseudogap $\Delta_{PG}$ which is prominent in the underdoped side. This leads to the following expression for the amplitude of the antinodal gap component: $\Delta_{AN}(T)=\sqrt{(\Delta_S(T))^2+(\Delta_{PG}(T))^2}$ where $\Delta_S(T)$ is considered as a BCS gap and therefore vanishes at $T_c$. 

 \par
From this equation it appears that only the $\Delta_{PG}$ term subsists at $T_c$. By extrapolating the location of the Raman antinodal peak at $T_c$ from the temperature dependence of the antinodal Raman response as a function of doping level (shown in the inset of Fig.~3), we can estimate $\Delta_{PG}(T=T_{c})$ over a wide doping range. The pseudogap component normalized to $T_c$  (full triangles) is plotted in Fig.~3. Below $p\approx0.12$, the pseudogap energy is hard to evaluate due to the weakness of antinodal Raman response intensity (see Fig.~1(d)) which is a consequence of the antinodal quasiparticle destruction induced by the pseudogap phase [\onlinecite{YG-PRB,LeTacon_NaturePhysics06,Shen_Science05, McElroy_PRL05}].
The doping dependences of the nodal (full circles) and antinodal (full squares) gap components normalized to $T_c$ are also reported. They have been respectively deduced from the $B_{2g}$ and $B_{1g}$ Raman spectra measured at $T=10 K$. 
\par
At first glance Fig.~3, reveals that the superconducting nodal component follows $T_c$ for all the doping levels with $2\Delta_N/k_{B}T_c\sim 6.4$ (Fig.~3). This contrasts with the antinodal superconducting component whose ratio departs from the nodal one and blows up close to optimal doping. 
The $\Delta_{AN}-\Delta_{N}$ splitting occurs when the pseudogap ratio measured at $T_{c}$ crosses the nodal one. 
ERS data do not allows us to directly probe the pseudogap ratio at low temperature ($T \ll T_c$). However we can infer from the $\Delta_{AN}-\Delta_{N}$ splitting that the pseudogap at $T=10~ K$ becomes significant below the optimal doping level. The pseudogap increases in magnitude and controls the antinodal component below $p\approx0.16$. In this regime the antinodal component is then progressively disconnected from superconductivity while the nodal one exhibits the same behavior throughout the superconducting dome.    

\par
In this picture the departure of the nodal component temperature dependence (shown in Fig.~2) from a BCS-like temperature dependence can be explained by a contribution of the pseudogap amplitude which would increase as the temperature is raised up to $T_{c}$. This last effect is suggested by the experimental data in the inset of Fig.~3 and recent ARPES data [\onlinecite{Lee_Nature07}]which reveal an increasing of the antinodal component as the temperature is raised in the underdoped regime. 

\par
In conclusion our experimental findings establish the nodal component as a good candidate for the  superconducting order parameter over a wide range of doping. The amplitude of the nodal gap shows a significant softening upon increasing temperature and scales with $T_c$ throughout the superconducting dome. By contrast the antinodal component of the gap is disconnected from superconductivity in the underdoped regime where it is largely controlled by the pseudogap amplitude. The physical origin of the pseudogap $\Delta_{PG}$ remains an important issue that we have to tackle in order to understand the mysterious evolution of the antinodal component as a function of doping level. 

\newpage
\begin{figure}
\begin{center}
\includegraphics[width=15cm]{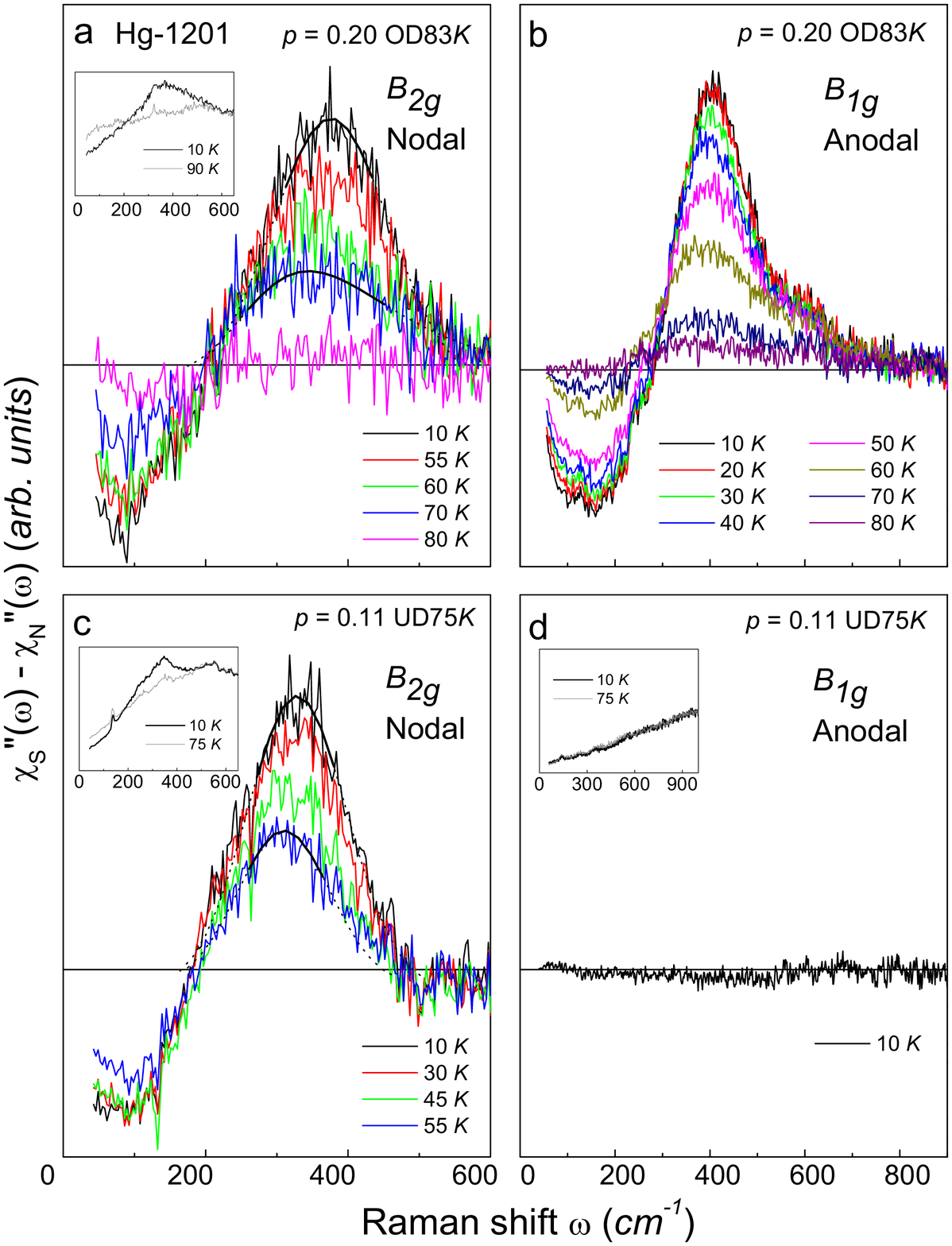}
\end{center}\vspace{-7mm}
\caption{(color online) Nodal ($B_{2g}$) and antinodal ($B_{1g}$) Raman response functions $\chi ^{\prime \prime}(\omega)$ substracted from the one just above $T_c$ as a function of temperature (up to $T_c$) for two doping levels: $p$ = 0.20 [(a) and (b)] and 0.11 [(c) and (d)]. The insets in pannels (a), (c) and (d) show the superconducting and normal Raman responses separetly. All the Raman responses have been obtained with the red ($1.9 eV$) excitation line. The doping value $p$ is inferred from $T_c$ using equation of Presland et al from Ref. [\onlinecite{PreslandPhysicaC91}]: $1-T_c/95.5 = 82.6\,(p-0.16)^2$. The dashed lines are guides for the eyes and track the locations of the superconducting peak maxima (in bold line).} 
\label{fig1}
\end{figure}

\begin{figure}
\begin{center}
\includegraphics[width=10cm]{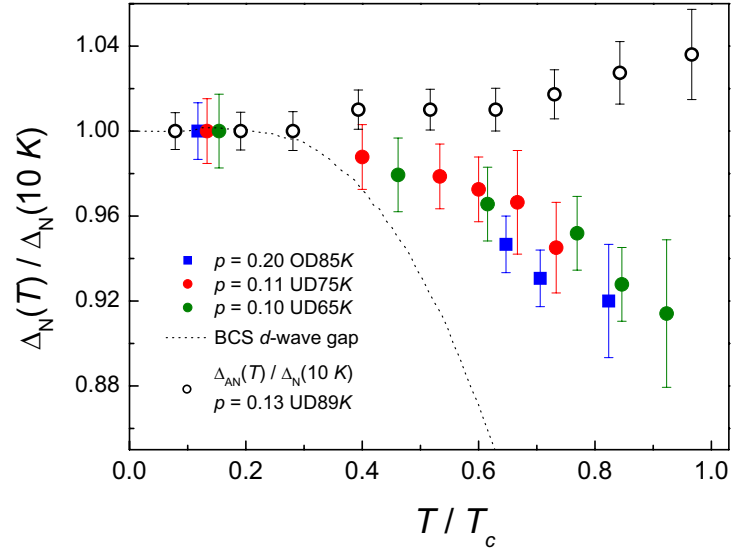}
\end{center}\vspace{-7mm}
\caption{(color online) Temperature dependences of the nodal superconducting component for several doping levels. $\Delta_{N}$ has been normalized to its value at $T = 10K$. $\Delta_{N}$ values have been determined from the locations of the peaks in the $B_{2g}$ Raman spectra of Hg-1201. The dashed line corresponds to the temperature dependence of a BCS $d$-wave gap from Ref. [\onlinecite{Carbotte_PRB95}]. For comparison the temperature dependence of the antinodal gap component for $p = 0.13$ is also shown.}

\label{fig2}
\end{figure}

\begin{figure}
\begin{center}
\includegraphics[width=10cm]{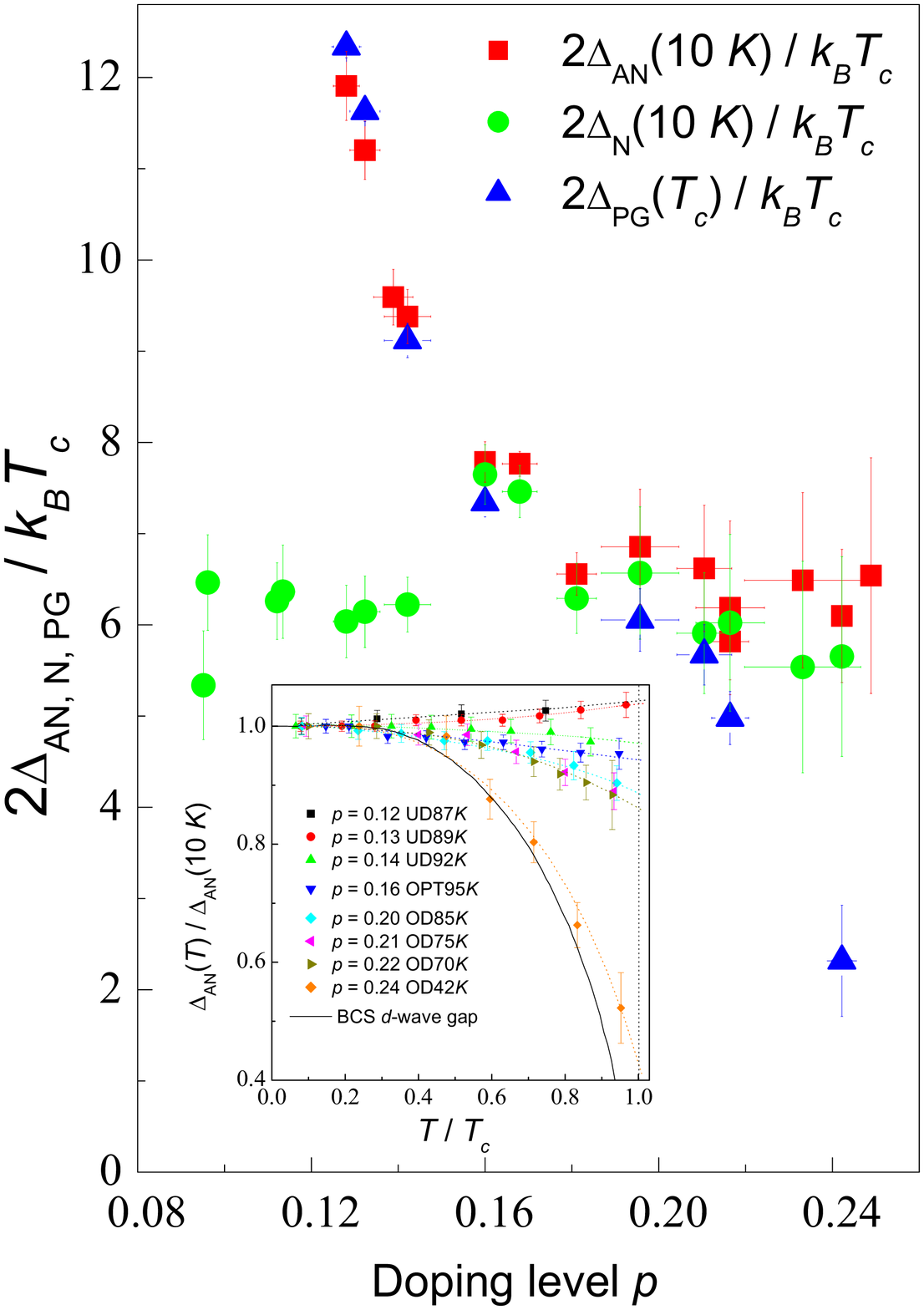}
\end{center}\vspace{-5mm}
\caption{(color online) Doping dependence of the  $2\Delta_{N}(T = 10~K)/k_{B}T_c$ and $2\Delta_{AN}(T = 10K)/k_{B}T_c$ ratios deduced from the locations of the nodal and antinodal gap components in the Raman spectra. The $2\Delta_{PG}/k_{B}T_c$  ratio is obtained from the extrapolation of the antinodal gap component at $T_{c}$ as shown in the inset where the temperature evolution of the normalized $\Delta_{AN}(T)$ is depicted [\onlinecite{Guyard_PRB77}]}.

\label{fig3}
\end{figure}

\section*{ACKNOWLEDGEMENTS}
We are grateful to A.~Georges, M.~Civelli, J.~L.~Tallon, Ph.~Monod, E.~Sherman, A.~J.~Millis and G.~Kotliar for very helpful discussions. Correspondences and requests for materials should be adressed to A.S (alain.sacuto@univ-paris-diderot.fr).


\begin{thebibliography}{9}
\bibitem{note1} We have performed Raman measurements up to $1200 ~cm^{-1}$without detecting any difference between the normal and superconducting states.
\end{thebibliography}
\end{document}